\documentclass[aps,prd,12pt,citeautoscript,superscriptaddress,showpacs,scrartcl]{revtex4-1}
\usepackage{xcolor}
\usepackage[colorlinks=true, pdfstartview=FitV,linkcolor=blue, citecolor=blue, urlcolor=blue]{hyperref}

\usepackage[bottom]{footmisc}

\usepackage{graphicx}

\makeatletter
\newcommand*\bigcdot{\mathpalette\bigcdot@{.5}}
\newcommand*\bigcdot@[2]{\mathbin{\vcenter{\hbox{\scalebox{#2}{$\m@th#1\bullet$}}}}}
\makeatother

\newcommand{\be}{\begin{equation}}
\newcommand{\ee}{\end{equation}}
\newcommand{\bea}{\begin{eqnarray}}
\newcommand{\eea}{\end{eqnarray}}

\newcommand{\mmu}{\mu}
\newcommand{\eeta}{\zeta}
\newcommand{\kkappa}{\alpha}
\newcommand{\h}{y}

\newcommand{\A}{{\rm a}}

\newcommand{\q}{{\cal H}^2}

\newcommand{\vareps}{\eeta}

\newcommand{\bl}{\textcolor{black}}

\usepackage{psfrag}
\usepackage{latexsym,array,theorem,mathrsfs,subfigure,bm,float}
\usepackage{amsmath}
\usepackage{amsfonts}
\usepackage{amssymb}

\begin{document}

\title{Anisotropy screening in  Horndeski cosmologies}






\author{Alexei~ A.~Starobinsky}
\email{\tt alstar@landau.ac.ru}
\affiliation{
L.D.Landau Institute for Theoretical Physics RAS, 
Moscow 119334, Russia}
\affiliation{
Department of General Relativity and Gravitation, Institute of Physics,\\
Kazan Federal University, 
420008 Kazan, Russia
}
\author{Sergey~V.~Sushkov}
\email{\tt sergey_sushkov@mail.ru}
\affiliation{
Department of General Relativity and Gravitation, Institute of Physics,\\
Kazan Federal University, 
420008 Kazan, Russia
}
\author{Mikhail~S.~Volkov}
\email{\tt volkov@lmpt.univ-tours.fr}
\affiliation{
Institut Denis Poisson, UMR - CNRS 7013, \\ 
Universit\'{e} de Tours, Parc de Grandmont, 
37200 Tours, France}
\affiliation{
Institute for Theoretical and Mathematical Physics, \\ 
Lomonosov Moscow State University,
Leninskie Gory, GSP-1, 119991 Moscow, Russia}
\affiliation{
Department of General Relativity and Gravitation, Institute of Physics,\\
Kazan Federal University, 
420008 Kazan, Russia
}

\ 

\vspace{1 cm}

\begin{abstract}

\vspace{1 cm}

We consider anisotropic cosmologies in a particular  shift-symmetric Horndeski theory containing the 
$G^{\mu\nu}\partial_\mu\phi \partial_\nu\phi$ coupling, where $G^{\mu\nu}$ is the Einstein tensor. 
This theory  admits stable in the future self-accelerating cosmologies 
whose tensor perturbations propagate with the velocity very close to the speed of light such that
the theory agrees with the 
gravity wave observations. Surprisingly, we find that the 
anisotropies within the Bianchi I homogeneous  spacetime model
are screened  at early time by the scalar charge, 
whereas at late times they are  damped  in the usual way. 
Therefore, contrary to what one would normally expect, 
\bl{the early state of the universe in the theory  
cannot be anisotropic and (locally) homogeneous in the absence of spatial curvature. 
The early universe cannot be isotropic  either, because it should  then be unstable 
with respect to inhomogeneous perturbations. As a result, the early universe should be inhomogeneous. 
At the same time, we find that in the spatially curved  Bianchi IX case the anisotropies can be strong 
at early times  even in the presence of a scalar charge. }

\end{abstract}


\maketitle

\section{Introduction} 

It is usually assumed that the state of the universe close to the initial singularity should be strongly 
anisotropic \cite{Belinsky:1970ew}, \cite{Collins:1972tf}, \cite{Belinsky:1982pk}. 
This belief  is based on the fact that spatial anisotropies produce in 
the Einstein equations terms   proportional to the inverse square of the volume, $1/V^2$,
which become dominant when one goes backwards in time. In other words, anisotropic perturbations 
grow to the past.  When the universe expands, the anisotropy  terms decrease faster  than  the 
contribution of other forms of energy subject to the dominant energy condition and the universe rapidly approaches 
a locally  isotropic state during inflation \cite{Starobinsky:1982mr}, \cite{Wald:1983ky}
(without the inflationary stage, this process may require a longtime or may not happen at all due to the 
possibility of a recollapse). 
Therefore, thinking about the early  history of the universe, one could expect 
the isotropic phase of inflation to be generically  preceded by an anisotropic 
phase. 
Although this argument seems quite robust, we shall present in what follows 
a peculiar cosmology  whose anisotropies are damped at early times, hence  the existence 
of a primary anisotropic phase is not as universal as one might  think. 

The theory we wish to discuss is the particular subset of the Horndeski theory 
for a gravitating scalar field 
\cite{Horndeski:1974wa} defined by the action \eqref{Fab5} below. 
Its homogeneous and isotropic cosmologies were studied in 
\cite{Sushkov:2009hk}, \cite{Starobinsky:2016kua}, but later it was discovered  that 
theories of this type should be  disfavored  
because they predict the speed of gravity waves (GW)  different from the speed of light 
\cite{Creminelli:2017sry,Ezquiaga:2017ekz,Baker:2017hug}, whereas 
the recent GW170817 event shows that the GW speed is equal to the speed of light with very high precision \cite{GW}. 
However, this constraint applies rather to some solutions of the theory than to the theory itself. 
The theory  admits stable in the future self-accelerating cosmologies 
whose tensor perturbations propagate with the velocity very close to the speed of light, the relative 
difference being proportional to $1/V$. Therefore, the theory can perfectly agree with the GW observation of \cite{GW}
at late times, and we can extrapolate it to the early times as well since no observational data  about the GW speed 
at redshifts $z>0.3$ are currently available. 

\bl{
We shall therefore study anisotropic cosmologies of the simplest Bianchi I  homogeneous spacetime type 
within this Horndeski model. 
Surprisingly, we find that the anisotropies are screened at early times by the scalar charge, hence 
 the standard argument in favor of strong anisotropies at early times does not always apply.
However, the universe cannot be isotropic in this limit either, since it would then be 
unstable with respect to inhomogeneous perturbations. This suggests that the early universe 
should be inhomogeneous. 
At the same time, our numerics suggest that  the universe can be strongly anisotropic 
close to the initial singularity within the Bianchi IX class, therefore the anisotropy screening is not generic for 
all Bianchi models.}

\section{Isotropic case}
\setcounter{equation}{0}
To begin with, we summarize the essential properties of the isotropic solutions, 
some of which have never been discussed  before. 
We consider the  theory 
\be                              \label{Fab5}
S=\frac12\int\left(\mmu\, R-(\kkappa\, G_{\mu\nu}
+\varepsilon\, g_{\mu\nu})\nabla^\mu\phi\nabla^\nu\phi-2\Lambda\right)\sqrt{-g}\, d^4x
\equiv\frac12 \int L\,d^4x\,, 
\ee
where 
$\mmu=M_{\rm Pl}^2$ is the Planck mass squared, 
\bl{
the parameter $\alpha$ has the  dimension of length squared, $[\alpha]=[L^{2}]=[M^{-2}]$ (we assume $c=\hbar=1$),
the parameter $\varepsilon$ is dimensionless, while $\Lambda$ is related to the cosmological constant $\bm{\Lambda}$ via 
$\Lambda=\mu{\bm{\Lambda}}$, one has $[\Lambda]=[L^{-4}]$.
We consider the theory (2.1) as a
classical theory of gravity valid for curvatures much less than
the Planck curvature. 
No lower energy or
curvature cutoff is required for its self-consistency, in
particular for the absence of ghosts.
}
Let us choose  the spacetime metric as 
\be                            \label{FLRW}
ds^2=-dt^2+\A^2(t)\left[
dx_1^2+dx_2^2+dx_3^2
\right], 
\ee
the case of more general homogeneous and isotropic metrics, including also an extra matter, 
was considered in \cite{Starobinsky:2016kua}. 
Assuming the scalar field $\phi$ to depend only on time, 
the Friedmann equation for the Hubble parameter $H=\dot{\A}/\A$  is (see \cite{Starobinsky:2016kua} 
for the explicit form of all equations in the theory)
\bea             \label{eq}
3\mmu\, H^2=
\frac12\,(\varepsilon-9\kkappa\, H^2)\,\dot{\phi}^2
+\Lambda. 
\eea
The equation for the scalar $\phi$ can be integrated once to give 
\bea                      \label{eq1}
\left(3\kkappa\, H^2-\varepsilon\right)\dot{\phi}=\frac{C}{\A^3}, 
\eea
where the integration 
constants $C$ is the scalar charge associated  with the invariance 
of the action under shifts  $\phi\to \phi+\phi_0$.  
\bl{
The notion of a scalar charge usually 
appears in the case of a complex scalar field, but we shall use it
here to denote the amount of the real scalar field $\phi$ according to
this definition.}

If $C=0$ then one has either 
\be                     \label{atr}
H^2=\frac{\Lambda}{3\mmu},~~~~\dot{\phi}=0,
\ee
or
\be                    \label{atr1}
H^2=\frac{\varepsilon}{3\kkappa},~~~~\dot{\phi}^2=\frac{\Lambda}{\varepsilon}-\frac{\mmu}{\kkappa},
\ee
in both cases the metric is pure de Sitter \bl{(all exact de Sitter
solutions with $\dot\phi\not= 0$ in the generic scalar-tensor theory  {without}
a derivative coupling of the scalar field to gravity can be found in  \cite{Motohashi:2019tyj}).}

If the charge $C$ does not vanish, then its effect should become 
negligible  for $a\to\infty$,
as seen from \eqref{eq1}, hence the solutions should approach either \eqref{atr} or \eqref{atr1} 
at late times (since $H^2$ and $\dot{\phi}^2$ should be positive, 
this imposes restrictions on values of the 
theory parameters.) 
If $C\neq 0$ then $\dot{\phi}$ can be algebraically expressed in terms of $\A,H$. 
Using  the values of the Hubble parameter and scale factor at present, $H_0,\A_0$, we
introduce dimensionless variables $y=(H/H_0)^2$, $a=\A/\A_0$ and $\psi=({3\kkappa H_0^2 \A_0^3}/C)\, \dot{\phi}$,
and also dimensionless parameters 
\be                    \label{par}
\Omega_0=\frac{\Lambda}{  3\mmu H_0^2   },~~~~~ 
\Omega_6=\frac{C^2}{18\,\kkappa\,\A_0^6\,H_0^4\,\mmu},~~~
\vareps=\frac{\varepsilon}{3\kkappa\,H_0^2}
\ee
\bl{(notice that the roman symbol ${\rm a}$ denotes the dimensionful scale factor, while 
$a$ stands for its dimensionless version).}
Equations \eqref{eq}, \eqref{eq1} then assume the form 
\be                          \label{HHH}
\h=\Omega_0
+\frac{\Omega_6 \left[\vareps-3\h \right]}{a^6 \left[\vareps-\h \right] ^2},~~~~
~~~~\psi=\frac{1}{ a^3(\h-\eeta)  }\,, 
\ee
and since they should hold if $y=a=1$, it follows that 
\be                          \label{HHH1}
\Omega_6=(\zeta-1)^2\frac{(1-\Omega_0)}{\zeta-3}. 
\ee
These equations determine $y(a)$ and $\psi(a)$, which determine $a(t)$ and $\phi(t)$. 

Before considering solutions of the equations, let us study conditions 
for their stability. 
Considering small fluctuations
$g_{\mu\nu}\to g_{\mu\nu}+\delta g_{\mu\nu}$, $\phi\to \phi+\delta\phi$, 
 the metric perturbation can be decomposed 
 into the scalar, vector, and tensor parts in the standard way \cite{weinberg}, while $\delta\phi$ 
 can be gauged away using the residual freedom of infinitesimal 
 reparametrizations of the time coordinate, hence $\delta\phi=0$. The second variation 
 of the action then splits into three independent parts describing the two tensor polarizations 
 and the scalar mode (the vector sector contains no dynamics). Each of these parts has the structure 
 \be                 \label{ITs}
 I=\frac{M_{\rm Pl}^2}{2}\, \int {\rm K}\left(  
\dot{X}^2-c_s^2\,\frac{P^2}{\A^2} \, X^2
   \right)\, \A^3\,d^4x\,,
 \ee
 where  $P$ is the spatial momentum. 
 In the case of tensor perturbations, $X$ is the tensor mode amplitude, while 
 the kinetic term and the sound speed squared are 
 \be                  \label{inT}
 {\rm K}_T=1+\Omega_6\psi^2,~~~~~~c_T^2=\frac{1-\Omega_6 \psi^2}{1+\Omega_6 \psi^2}\,,
 \ee
 with $\psi$ given by \eqref{HHH}. 
In the scalar sector one has  $X=\delta g_{00}$ while
\bl{
\bea                       \label{inS}
 {\rm K}_S&=&\frac{[\Omega_6+a^6(y-\zeta)^2]\,[\Omega_6\,(3y+\zeta)-a^6\,(y-\zeta)^3]}
 {a^6 \,y\,(y-\zeta)^2\, [3\,\Omega_6+a^6\,(y-\zeta)^2)]^2}\,, \nonumber \\
 c_S^2&=&\frac{[3\,\Omega_6+a^6\,(y-\zeta)^2]\, \{ \left[\,a^6\,(y-\zeta)^3+(\Omega_6/3)(13y-3\zeta)\,\right]^2-(52/9)\,\Omega_6^2\, y^2   \}
 }{[\Omega_6+a^6\,(y-\zeta)^2]\,[\Omega_6\,(3y+\zeta)-a^6\,(y-\zeta)^3]^2}\,. 
 \eea
 }
 The functions ${\rm K}_S$, ${\rm K}_T$, $c_S^2$, $c_T^2$ should be positive for the background solutions to be stable.

We can now analyze  solutions of \eqref{HHH}, and the simplest way to get them is to transform the first equation in
\eqref{HHH} to
\be                 \label{a6}
a^6=\frac{\Omega_6(\zeta-3y)}{(y-\zeta)^2(y-\Omega_0)}. 
\ee
Therefore, if $a\to 0$ then $y\to \zeta/3$, while if $a\to \infty$ then either $y\to\Omega_0$ or $y\to \zeta$.  
\bl{
In general, this defines three  different solutions $y(a)$, but only one of them extends to the whole interval $a\in[0,\infty)$, 
the two others being unphysical \cite{Starobinsky:2016kua}. 
One has to have $\zeta\sim\varepsilon/\alpha>0$, 
since otherwise $y(a)$ is not positive definite, yielding  
solutions with ghost \cite{Starobinsky:2016kua}. 
Therefore, the parameters $\varepsilon$ and $\alpha$ should have the same sign.
Let us  assume first that $\alpha>0$ and $\varepsilon>0$, 
hence  $\Omega_6>0$. }

It is natural to assume that $0<\Omega_0<1$. Then the positivity of  $\Omega_6$ 
defined by \eqref{HHH1}  requires that $\zeta>3$.  In this case there exists  only one solution $y(a)$ of \eqref{a6}, 
which fulfills 
\be                       \label{sol}
\frac{\zeta}{3}\leftarrow y \rightarrow \Omega_0~~~~~~\mbox{as}~~~~~0\leftarrow a\rightarrow \infty. 
\ee
A direct verification reveals that ${\rm K}_T>0$ and ${\rm K}_S>0$ everywhere for this solution, hence the ghost is absent, 
while at large $a$ one has $c^2_T>0$ and  $c^2_S>0$, hence the solution 
 is free in this limit also from gradient instabilities. 
The profile of this solution is shown in Fig.1. 

\bl{The solution has two inflationary stages: an early 
inflation driven by the scalar $\phi$,  with the Hubble rate
 $H^2=H_0^2\,y\approx H_0^2\,\zeta/3=\varepsilon /(9\alpha)\equiv H_{e}^2$, 
and a  late inflation driven by the cosmological constant, with 
$H^2\approx H_0^2\,\Omega_0=\Lambda/(3\mu)={\bm \Lambda}/3\equiv H_{l}^2$.  
Since $H_{l}$ is small, 
to have a hierarchy between the two inflationary scales, $H_l\ll H_e$, one should
assume  that $H_{\rm e}\sim \varepsilon/\alpha\sim \zeta$ is large, hence the 
coefficient  $\kkappa$ in \eqref{Fab5} should be small. At the same time, one needs 
$H_{e}^2  \ll M_{\rm Pl}^2=\mu$ for the classical theory to apply, hence $\alpha\gg 1/\mu\equiv L_{\rm Pl}^2$.
Therefore, even though $\alpha$ is small, it should be much larger than the Planck length squared. 
Similarly, although  $\zeta$ is large, there is an upper bound 
$\zeta=\varepsilon/(3\alpha H_0^2)\ll \mu/H_0^2\approx 10^{122}$. 
 }
 
 \bl{One should say that, although the early inflationary stage is regular
in terms of geometry, the scalar field is singular since one has 
$\psi\propto a^{-3}\propto e^{-3H_{e}t}$ for $a\to 0$ and $t\to -\infty$.
As a result, this  is a kind of  a ``fast-roll" inflation, which  makes
questionable its usage for constructing viable cosmological models with
approximately flat (scalefree) spectra of initial scalar and tensor
perturbations generated by quantum-gravitational effects. 
At the same time, as we shall see shortly, the universe should become inhomogeneous 
at early times, and that may change the way the spectra are derived. 
However, studying these 
issues  goes beyond the scope of our purely classical analysis.
 }

\begin{figure}[h]
\hbox to \linewidth{ \hss

	
				\resizebox{9cm}{7cm}{\includegraphics{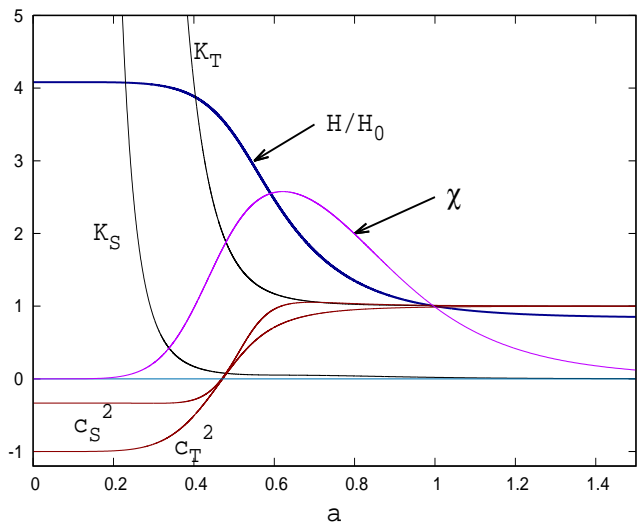}}

	
\hspace{1mm}
\hss}
\caption{
Profile  of $\sqrt{y}=H/H_0$ for $\Omega_0=0.7$ and $\zeta=50$. 
The kinetic terms ${\rm K}_S$ and ${\rm K}_T$ are always positive hence 
the ghost is absent.  
One has ${\rm K}_T\to 1$ while ${\rm K}_S\propto 1/(\zeta-\Omega_0)\Omega_0 a^6) $ at large $a$. 
The sound speeds  ${\rm c}_S^2$ and ${\rm c}_T^2$ 
approach unity at late times but become negative at small $a$,  showing 
gradient instabilities  with respect to inhomogeneous perturbations. 
The amplitude $\chi$ is the anisotropy 
defined by \eqref{chi} (assuming that  ${\cal B}=1$). 
}
 \label{Fig1}
\end{figure}

As $a\to\infty$ one has ${\rm K}_T=1+{\cal O}\left(1/a^6\right)$
 hence the GW speed approaches the speed of light.  At the same time, 
 the present time moment corresponds to finite values $y=a=1$, when 
  \be
{\rm K}_{T}-1=\frac{1-\Omega_0}{\zeta-3}. 
\ee
Since $\zeta$ and $\Omega_0$ determine the Hubble rates of the early and late inflations, 
the ratio $\zeta/\Omega_0$ should be of the order of the  early  inflation energy scale divided by the late inflation 
energy scale, in which case \bl{ ${\rm K}_{T}-1\sim \Omega_0/\zeta\sim (H_l/H_e)^2 \ll 10^{-15}$.  }
This agrees  with the observed bound  on the relative difference 
 between the GW velocity and the speed of light \cite{GW}. 

\bl{
The solution develops gradient instabilities for small $a$ when 
 $c_T^2$ and $c_S^2$ become 
negative because, as seen from \eqref{inT} and \eqref{inS},  $c_T^2\to -1$ and $c_S^2\to-1/3$ as $a\to 0$. 
If $\zeta\gg 1$ then zeroes $c_T^2$ and $c_S^2$  are 
close to the point where $y=\zeta /4$. 
Before discussing 
these instabilities, we shall consider below another property 
of the solution -- the screening  of anisotropies. }

\bl{
To finish this section, there are solutions with two inflationary stages also if $\alpha<0$, $\varepsilon<0$  and $\Omega_6<0$. 
However, in this case one has  $0<\zeta<3$, and since $\zeta$ cannot be large, the difference $K_T-1$ is not small, 
hence these solutions give a wrong value  for the GW speed. }

\section{Bianchi I  -- anisotropy screening}
\setcounter{equation}{0}

Let us consider the Bianchi I metric
\be                        \label{ma}
ds^2=-N^2\,dt^2+\A_1^2\,dx_1^2+\A_2^2\,dx_2^2+\A_3^2\,dx_3^2\,,
\ee
where $N,\A_k$ and also $\phi$ are functions of $t$. 
Injecting to  \eqref{Fab5}, 
the Lagrangian  is 
\be                    \label{LL}
L=-3\A^3\left(\frac{2\mmu}{N}
+\frac{\kkappa\, \dot{\phi}^2}{N^3}\right){\cal H}^2
+\left(\frac{\varepsilon \, \dot{\phi}^2}{N}-2N\Lambda\right)\,\A_1\, \A_2\, \A_3\,,
\ee
with 
\be                          \label{Q}
3\A^3 \q\equiv \A_1\,\dot{\A}_2\,\dot{\A}_3
+\A_2\,\dot{\A}_1\,\dot{\A}_3
+\A_3\,\dot{\A}_1\,\dot{\A}_2=
3\A^3 \left(\frac{\dot{\A}^2}{\A^2}-\dot{\beta}_{+}^2-\dot{\beta}_{-}^2\right),
\ee
where
$
\A_1=\A\, e^{\beta_{+}+\sqrt{3}\beta_{-}}$, 
$\A_2=\A\,e^{\beta_{+}-\sqrt{3}\beta_{-}}$,
$\A_3=\A\, e^{-2\beta_{+}}$. 
The field equations can be obtained by varying $L$  with respect to $N,\beta_\pm,\phi$ and then 
setting $N=1$. This yields 
\bea             \label{eq3}
3\mmu\, {\cal H}^2=
\frac12\,(\varepsilon-9\kkappa\, {\cal H}^2)\,\dot{\phi}^2
+\Lambda, \\
\left(\sigma \A^3\dot{\beta}_\pm\right)^{\mbox{.}}=0, \label{3}  \\
\A^3\left(3\kkappa\, {\cal H}^2-\varepsilon\right)\dot{\phi}=C,  \label{eq4}
\eea
with  $\sigma=2\mmu+\kkappa\dot{\phi}^2$, where $C$ is the scalar charge. 
If the anisotropies $\beta_\pm$ vanish, these equations reduce to \eqref{eq},\eqref{eq1}. If  anisotropies
do not vanish, then one has from \eqref{3}
\be             \label{XXX}
\dot{\beta}_{\pm}=2\mmu\,\frac{{\cal B}_{\pm}}{ \sigma \A^3 }, 
\ee
where ${\cal B}_\pm$ are  integration constants. Let us see what this implies first in the case when 
the scalar charge is zero, $C=0$. Then \eqref{eq4} can be solved by 
 $\dot{\phi}=0$ while \eqref{eq3} and \eqref{XXX} yield 
 \be
 \frac{\dot{\A}^2}{\A^2}=\dot{\beta}_{+}^2+\dot{\beta}_{-}^2+\frac{\Lambda}{3\mmu},~~~~~
 \dot{\beta}_{\pm}=\frac{{\cal B}_{\pm}}{ \A^3 }.
 \ee
The  anisotropy terms on the right in the first equation decay with time, hence  anisotropie contribution 
becomes irrelevant and the universe rapidly approaches the isotropic de Sitter phase \eqref{atr}. 
However, the  anisotropy terms become dominant at small $\A$, 
when  one can neglect the $\Lambda$-term and the universe is described by the Kasner metric
for which $\A_k\propto t^{s_k}$ where the exponents $p_k$ are expressed in terms of ${\cal B}_\pm$
and fulfill $p_1+p_2+p_3=p_1^2+p_2^2+p_3^2=1$. 
This supports the standard 
view according to which anisotropies should be important close to the initial singularity. 
 
 If $C=0$ then \eqref{eq4} can be solved also by setting 
 $
 3\kkappa \q-\varepsilon=0,
 $
 which gives 
\be
 \frac{\dot{\A}^2}{\A^2}= \dot{\beta}_{+}^2+\dot{\beta}_{-}^2+\frac{\varepsilon}{3\kkappa},~~~~~~~
 \dot{\beta}_{\pm}=\frac{2\mmu\varepsilon\, {\cal B}_{\pm}}{( \mmu\varepsilon+\kkappa\Lambda )\,\A^3}, ~~~~~~~
 \dot{\phi}^2=\frac{\Lambda}{\varepsilon}-\frac{\mmu}{\kkappa}. 
 \ee
 The solution approaches the isotropic  de Sitter phase \eqref{atr1} at late times, 
 but at early times  the anisotropies are again dominant. 
 \bl{However, unless if $\alpha\Lambda=\epsilon\mu$, this solution is unphysical because 
 $\dot{\phi}\sim \psi$ approaches at late times a constant nonzero value, therefore, according to \eqref{inT}, 
 the GW velocity is not equal to the speed of light. 
 }
 
 Assume now that the scalar charge does not vanish, $C\neq 0$. Then one obtains from \eqref{eq4} 
  \be
 \dot{\phi}=\frac{C}{\A^3(3\kkappa\q-\varepsilon)}. 
 \ee
 Injecting this  to \eqref{eq3}, 
 setting 
 ${\cal H}^2={H}_0^2\, y$ 
and introducing the same  $a,\psi,\Omega_0,\Omega_6,\zeta$ as above,
 one obtains exactly the same equations as in \eqref{HHH}. 
 Their solution  for $y(a)$ and $\psi(a)$ is the same as the one described above and 
 shown in Fig.1. This time, however, 
 it describes a Bianchi I  spacetime  with 
 the anisotropies expressed by \eqref{XXX}, 
\be                  \label{beta}
\dot{\beta}_\pm=\frac{{\cal B}_\pm}{\A^3(1+\Omega_6\psi^2)}\,,
\ee
and with 
 the Hubble rate 
\be               \label{Hub}
\frac{\dot{\A}^2}{\A^2}=\dot{\beta}_{+}^2+\dot{\beta}_{-}^2+{\cal H}^2=
\left(\frac{\dot{\beta}_{+}^2+\dot{\beta}_{-}^2}{{\cal H}^2}+1\right){\cal H}^2\equiv 
H_0^2\left( \chi+ 1 \right)y. 
\ee
Here 
\be                  \label{chi}
\chi=\frac{{\cal B} }{a^6(1+\Omega_6\psi^2)^2 y}
\ee
is the relative contribution of the anisotropies to the total energy balance, its amplitude is 
${\cal B}=({\cal B}_{+}^2+{\cal B}_{-}^2)/( H_0a^3_0)^2$. 
Since the universe 
is highly isotropic at present, when $y=a=1$, one should assume that ${\cal B}\ll 1$, but this
does not mean that isotropies have always been small. 

Notice however that, according to \eqref{sol}, one has at early times 
\be              \label{late}
\h\approx \frac{\eeta}{3},~~~~~
\psi=\frac{1}{a^3(y-\zeta)}\propto a^{-3}~~~\Rightarrow~~~ a^3(1+\Omega_6\psi^2)\propto a^{-3}
\ee
and hence 
\be
\dot{\beta}_\pm\propto a^3~~~~\Rightarrow~~~~~ \dot{\beta}_{+}^2+\dot{\beta}_{-}^2\propto a^6. 
\ee
As a result, the anisotropies tend to zero for $a\to 0$ and their contribution to the total energy balance 
is $\propto a^6$ instead of $\propto 1/a^6$. Therefore,  the anisotropy effect is totally negligible at early times. 
This is true if only the scalar charge $C$ is nonzero, hence one can say that 
 anisotropies are ``screened by the scalar charge".  
Of course, the anisotropy contribution is suppressed at late time as well by the factor $1/a^6$ in \eqref{Hub}
(since $\psi\to 0$ as $a\to\infty$). As seen in Fig.\ref{Fig1}, 
the anisotropy $\chi$ defined by \eqref{chi} 
approaches zero at early and late times. 

\bl{
As there is no reason to assume  the scalar charge to be zero, it follows that 
the anisotropies are screened at early times in our theory,  at least in the Bianchi I case. 
This means that, unlike what one would  normally expect, 
the state of the universe close to the singularity cannot be \bl{anisotropic and homogeneous}. 
The early universe cannot be homogeneous and isotropic either,
because then it would develop at small $a$  the gradient instability  when the sound speed squared $c^2$ 
becomes negative, as seen in Fig.\ref{Fig1}. A gradient (not ghost) instability indicates that the universe
has a tendency to evolve to a different state. 
The instability is  present both 
in the tensor and scalar sectors and it exists only for inhomogeneous perturbations, 
since the corresponding potential term in the effective action \eqref{ITs} 
contains the factor of $P^2$. In fact, the same condition that insures the anisotropy damping, 
$\Omega_6\psi^2\gg 1$, guarantees  that $c_T^2$ and $c_S^2$ in \eqref{inT} and \eqref{inS} 
be negative. As a result, 
the early universe cannot be homogeneous and anisotropic, neither can it remain homogeneous and isotropic,
hence  it should evolve toward an inhomogeneous state. 
 }

\bl{
The latter conclusion applies in fact to the whole stage of the primary inflation. As discussed above, 
when moving  backward in time, the variable $y$ grows and the gradient instability starts at $y\approx \zeta/4$, 
while the inflation starts only at $y\approx \zeta/3$. 
Therefore, the primary inflation  falls  entirely within  the instability region, hence it should be 
inhomogeneous. The consideration of  an inhomogeneous inflation  
 is beyond the scope of the present paper,
but one may think that the spacetime geometry could then perhaps be described by something similar 
to the Gowdy metrics \cite{Gowdy:1973mu}. One might conjecture that 
 the homogeneous component of the scalar charge will then be 
 dispersed into its small-scale inhomogeneous fluctuations, 
so that the total spatially averaged value of $\Omega_6\psi^2$ does not exceed unity. 
However, an additional analysis is needed to study these issues.}

\bl{
One should also say that Eqs. \eqref{inT}, \eqref{inS} describing the perturbations 
have been derived for the homogeneous and isotropic backgrounds. They 
can be used also in the anisotropic Bianchi I case at early times, since the anisotropies 
are then damped. However,  at  the intermediate times,
when the anisotropies are not necessarily small, one should separately carry out the 
analysis of perturbations  and rederive the coefficients $K_T, K_S, c^2_T, c^2_S$ 
by taking the anisotropies into account. 
This would probably give a different value of the GW speed $c^2_T$ 
(for example, taking  the background inhomogeneities into account changes the GW speed \cite{Copeland}). 
However, the observational constraint on the GW speed apply only 
for relatively recent times,  when the anisotropies should be  small again, hence  one can 
use in this case Eqs. \eqref{inT}, \eqref{inS}. 
}

\section{Bianchi IX  case}
\setcounter{equation}{0}
One may wonder if the anisotropy screening  is typical only for the Bianchi I class or it occurs 
also for other Bianchi types. We shall therefore analyse  the Bianchi IX class, in which case 
the spacetime metric is 
\be                        \label{maa}
ds^2=-N^2\,dt^2
+\frac14\left(\A_1^2\,\omega_1\otimes \omega_1
+\A_2^2\,\omega_2\otimes \omega_2
+\A_3^2\,\omega_3\otimes \omega_3\right),
\ee
where $\omega_a$ are the invariant forms on $S^3$ subject to 
$
d\omega_a+\epsilon_{abc}\,\omega_b\wedge \omega_c=0,$
while $\A_k$ and the scalar $\phi$ depends only on time. 
The Lagrangian in \eqref{LL} generalizes to 
\be
8L=\frac{6\mmu\A^3 }{N}\left(\frac{{\cal K}N^2 }{\A^2}-{\cal H}^2  \right)
-\frac{3\kkappa\A^3}{N^3}\,\dot{\phi}^2\left({\cal H}^2+\frac{{\cal K}N^2 }{\A^2} \right)
+\left(\frac{\varepsilon}{N}\,\dot{\phi}^2-2N\Lambda \right)\A^3\,,
\ee
where ${\cal H}$ is the same as in \eqref{Q},
 while the anisotropy potential is 
\be                        \label{K}
{\cal K}=-\frac13\,e^{-8\beta_{+}}
\left(4e^{6\beta_{+}}\cosh^2(\sqrt{3}\beta_{-})-1\right)
\left(4e^{6\beta_{+}}\sinh^2(\sqrt{3}\beta_{-})-1\right).
\ee
Varying the Lagrangian and setting $N=1$ gives the 
equations  (with $\sigma=2\mmu+\alpha\dot{\phi}^2$)
\bea             \label{K1}
3\mmu\left({\cal H}^2+\frac{\cal K}{\A^2}\right)
+\frac{3}{2}\,\kkappa\,\dot{\phi}^2\left(3{\cal H}^2+\frac{\cal K}{\A^2}\right)= 
\frac{\varepsilon}{2}\,\dot{\phi}^2
+\Lambda,           \\                 
\left.\left.\frac{1}{\A^2}\right(\sigma \A\dot{\A}\right)^{\mbox{.}}
=\sigma\left(\frac12 {\cal H}^2
-\dot{\beta}^2_{+}
-\dot{\beta}^2_{-}\right)
+\left(\frac{\kkappa}{2}\,\dot{\phi}^2-\mmu\right)\frac{\cal K}{\A^2}
-\frac{\varepsilon}{2}\,\dot{\phi}^2+\Lambda,           \label{K2}    \\
\left(\sigma\A^3 \dot{\beta}_\pm \right)^{\mbox{.}}=
\A\left(\mmu-\frac{\kkappa}{2}\,\dot{\phi}^2\right)\frac{\partial{\cal K}}{\partial \beta_\pm},        \label{K3}   \\
\A^3\left(3\kkappa\,\left({\cal H}^2+\frac{\cal K}{\A^2}\right)-\varepsilon\right)\dot{\phi}=C,    \label{K4} 
\eea
where the first equation \eqref{K1} is actually the first integral of the remaining \eqref{K2}--\eqref{K4}. 
The effect of anisotropies is encoded in Eqs.\eqref{K1},\eqref{K4} only through the term 
${\cal K}\geq 1$. Therefore, applying the same transformations as before, 
one obtains instead of \eqref{HHH}  the equations 
\be                          \label{HHH2}
\h=\Omega_0+\frac{\Omega_2}{a^2}
+\frac{\Omega_6 \left[\vareps-3\h +\Omega_2/a^2\right]}{a^6 \left[\vareps-\h +\Omega_2/a^2\right] ^2},~~~~
~~~~\psi=\frac{1}{ a[a^2(\eeta-\h) +\Omega_2] }\,. 
\ee
\bl{ 
Although they resemble Eqs.\eqref{HHH}, they do not form a closed system since they 
contain  $\Omega_2=-{\cal K}/(H_0^2 \A_0^2)$  where ${\cal K}$ is the  function of the anisotropies 
defined by \eqref{K}. 
However, the system of equations becomes closed  in the isotropic case, when 
one can consistently set  $\beta_{\pm}=0$, which yields   ${\cal K}=1$ and $\Omega_2=-1/(H_0^2 \A_0^2)$. 
The solutions for $y(a)$ then can be obtained from \eqref{HHH2} by applying  the Cardano formula, 
and they are such that $y(a)$ approaches $\Omega_0$  as $a\to\infty$
but vanishes at $a=a_{\rm min}>0$ and becomes negative for $a<a_{\rm min}$ \cite{Starobinsky:2016kua}. 
These  solutions  describe bouncing universes which shrink from infinity  to 
the minimal  size $a_{\rm min}$ when the Hubble parameter $H=H_0\sqrt{y}$ vanishes,  
and then expand again \cite{Starobinsky:2016kua}. Such a bouncing behavior is due to the positive spatial curvature. 
These bounces exist for any however small value of $\Omega_0>0$. 
If $C=0$  then $\Omega_6=0$ and \eqref{HHH} reduce to 
\be
\h=\Omega_0+\frac{\Omega_2}{a^2},~~~~~\psi=\frac{1}{ (\eeta-\Omega_0)a^3 }\,, 
\ee
and using $y=(\dot{a}/a)^2/H_0^2$ yields 
\be
a(t)=\sqrt{\frac{3\mu}{\Lambda\, \A_0^2}}\cosh\left( \sqrt{ \frac{\Lambda}{3\mu}}\,t \right). 
\ee
This describes the de Sitter metric expressed in  coordinates with compact spatial sections.  }

These bounces can be generalized to include small anisotropies, because expanding the equations \eqref{K3}
for $\beta_{\pm}$ up 
to the fist order yields  $\left(\sigma\A^3 \dot{\beta}_\pm \right)^{\mbox{.}}=0$ 
and hence $\dot{\beta}_\pm=2\mu{\cal B}_\pm/(\sigma \A^3)$. 
Since the value of $\sigma \A^3$ is bounded below for a bounce, it follows that if the integration 
constants ${\cal B}_\pm$ are small, 
the anisotropies always  remain small and only produce a small correction to the Hubble rate. The  
zero of $H$ shifts  slightly due to 
the anisotropies, 
since one has 
\be
H^2=\dot{\beta}_{+}^2+\dot{\beta}_{-}^2+H_0^2\, y. 
\ee
Therefore, if $\beta_{\pm}=0$ then $H$ vanishes at $a_{\rm min}$ where $y$ vanishes, 
but if  $\dot{\beta}_\pm\neq 0$  then the zero of $H$ shifts to the region $a<a_{\rm min}$ where $y<0$.

Now, if $C=\dot{\phi}=0$ then the equations also admit the slightly anisotropic bounces, 
but they admit as well  strongly anisotropic solutions with initial singularity. In other words, increasing the 
amplitude of anisotropies shifts the bounce position more and more until it reaches $a=0$, after which 
the solutions are no longer bounces and show the initial curvature singularity. 
Equation \eqref{K1} then reduces to 
\be
\frac{\dot{\A^2}}{\A^2}=\dot{\beta}_{-}^2+\dot{\beta}_{+}^2-\frac{\cal K}{\A^2} +\frac{\Lambda}{3\mmu}\,,
\ee
where the positive anisotropy terms on the right are large enough to overcome the  negative  term, 
thereby eliminating 
the bouncing behavior. The 
solutions are then characterized by a sequence of  ``Kasner epochs" during which 
${\partial{\cal K}}/{\partial \beta_\pm}\approx 0$ 
 and the universe is approximately  described by the Kasner metric with 
 $\A^3 \dot{\beta}_\pm\equiv  {\cal B}_\pm\approx const.$
 The term ${\partial{\cal K}}/{\partial \beta_\pm}$ 
becomes important only during short moments  when ${\cal B}_\pm$ change, 
after which the next Kasner epoch with new values of ${\cal B}_\pm$  starts  \cite{Belinsky:1970ew}. 

One can wonder if a similar evolution near singularity is possible also when the scalar 
charge $C$ does not vanish ?
It seems at  first  that the answer should be negative, since  the Kasner epochs 
during which  the solution is approximately Bianchi I seem to be forbidden by the 
anisotropy screening.  

To clarify the situation, 
we  solved numerically the system of second order equations \eqref{K2},\eqref{K3} together with the  equation 
obtained by differentiating \eqref{K4}. The first order equation \eqref{K1} was used to constraint the initial values,
and we checked that the constraint propagates. We also checked that the scalar charge defined 
by \eqref{K4} remains constant during the evolution, as it should. We chose the initial data to describe  a slightly 
anisotropic  universe of a finite size and then integrated the equations to the past. It turns out that if the initial anisotropy 
is small, then the scale factor $\A(t)$ first decreases to the past, then passes through a minimal nonzero value, and then  
stars increasing. The solution is of the bounce type and the anisotropies always remain small. 
However, if the initial anisotropy is large enough, then the scale factor always decreases to the past
while the anisotropies {\it grow}. The singularity is strongly anisotropic.

\begin{figure}[h]
\hbox to \linewidth{ \hss

	
				\resizebox{9cm}{7cm}{\includegraphics{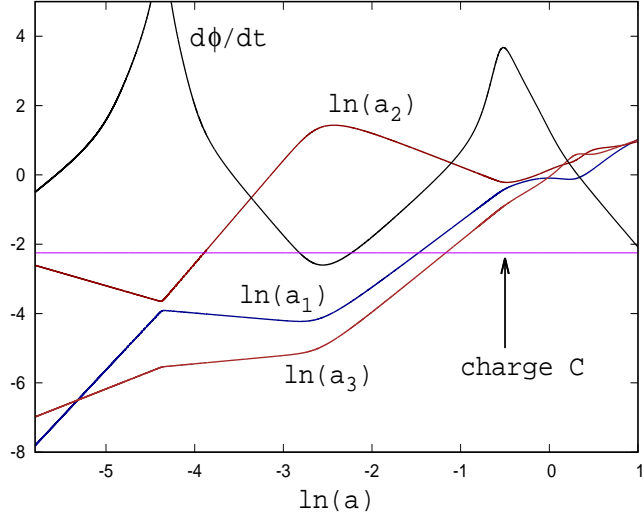}}

	
\hspace{1mm}
\hss}
\caption{Solution of Eqs.\eqref{K1}--\eqref{K4} for $\mmu=\Lambda=\varepsilon=1$,
$\kkappa=0.1$ and with the  initial values $\A=5$, $\dot{\beta}_{\pm}=0$, ${\beta}_{+}=0$, 
${\beta}_{-}=0.05$, $\dot{\phi}=0.02$ (assuming all parameters to be dimensionless and expressed 
in Planck units). 
The solution 
shows a sequence of Kasner epochs during which $\ln(\A_k)/\ln(\A)\approx const.$ 
The scalar $\dot{\phi}$ oscillates but the charge $C$ is constant. 
}
 \label{Fig2}
\end{figure}

A typical solution is shown in Fig.\ref{Fig2}. Surprisingly, it demonstrates a sequence of Kasner epochs
during which it must approach the Bianchi I regime. 
This seems to contradict the fact that the anisotropies should then be screened. 
However, the explanation  is
the following. 
In the Bianchi I case the anisotropies are screened  because $\psi\propto a^{-3}$, which makes large 
the denominator in \eqref{beta}. In the Bianchi IX case Eq.\eqref{HHH2} yields 
$y=\Omega_2/(3a^2)+{\cal O}(1)$ and $\psi=3/(2\,\Omega_2 \,a)+{\cal O}(1)$ for small $a$. 
Since $\Omega_2\sim {\cal K}$ one has $\psi\propto 1/({\cal  K} a)$, and since ${\cal K}\geq 1$,
it follows that $\psi$ grows {not faster} than $a^{-1}$. 
This implies that the denominator in \eqref{beta} behaves as $a/{\cal K}^2$ and 
tends to zero, hence  $\dot{\beta}_\pm$ expressed by  \eqref{beta} are large. 
Therefore, 
the anisotropies are not screened  in the Bianchi IX case, hence  their screening 
is not a generic feature  for  all Bianchi models. A more detailed analysis is needed to find out 
if the solutions can be chaotic \cite{Belinsky:1970ew}. 
\bl{The stability  of these  solutions with respect to inhomogeneous perturbations 
remains an open issue to study. }

\section{Conclusion}
\setcounter{equation}{0}

We studied anisotropic cosmologies  in the shift-symmetric, 
nonminimally coupled Horndeski model  \eqref{Fab5}. Even 
though this model is thought to be disfavored by the GW observations, its homogeneous and isotropic solution 
 propagates tensor perturbations with the velocity  that can be insensitively 
close to the speed of light. Surprisingly, 
it turns out that the spatial anisotropies in this theory 
get damped  at early times in the Bianchi I case, instead of being amplified.  
Therefore, the standard argument in favor of strong anisotropies at early times does not always apply.
However, it seems that the anisotropy screening is not generic for all Bianchi types, since 
our numerics suggest that  the universe can be strongly anisotropic 
close to the initial singularity within the Bianchi IX class, 
\bl{where positive spatial curvature is present. The scale factor
$a(t)$ and the scalar field $\phi(t)$ can also show a regular bouncing behavior  in
this case, similarly to the minimally coupled  scalar  field in the closed
Friedmann-Lema$\hat{\i}$tre-Robertson-Walker space-time~\cite{Starobinsky78}, but the  generic solution contains a curvature singularity.}

\bl{Although the anisotropies are screened at early times in the Bianchi I case, the universe does not approach 
an homogeneous and isotropic state, since it would then be unstable  with respect to inhomogeneous perturbations. 
This suggests that  the early stage of the universe  should be essentially inhomogeneous. To study such inhomogeneous 
cosmologies requires a separate analysis. }

To the best of our knowledge, a similar systematic analysis of anisotropic cosmologies for generic Horndeski models 
has never been carried out before, although anisotropic cosmologies with scalars have  been studied.
For example, anisotropies in the theory with a conformally coupled scalar field
\cite{Kamenshchik:2017ojc} and also in the $R+R^2$ gravity 
 \cite{Gurovich:1979xg}, \cite{Muller:2017nxg} have been discussed,
  both cases being conformally dual to the ordinary gravity with 
a scalar field. But in our case the theory cannot be conformally transformed to the Einstein frame. 
As a result, qualitatively new behavior of anisotropy takes place.

{\bf Acknowledgments:}  The work of A.A.S. and S.V.S was supported by the 
Russian Foundation for Basic Research, 
grant No.19-52-15008. 
The work of  M.S.V. was partly supported by the  French National Center of Scientific Research
within the joint French-Russian research program, Grant No. 289860, 
as well as by the Russian Government Program of Competitive Growth 
of the Kazan Federal University.



\providecommand{\href}[2]{#2}\begingroup\raggedright\endgroup

\end{document}